\documentclass[notitlepage,english,aps,floats,onecolumn,showpacs,nofootinbib,floatfix]{revtex4-2}
\usepackage{pslatex}
\usepackage[T1]{fontenc}
\usepackage[latin1]{inputenc}
\usepackage{graphicx}
\usepackage{epsfig}
\usepackage{longtable}
\usepackage{float}
\usepackage{calc}
\usepackage{ifthen}
\usepackage{amsmath}
\usepackage{hyperref}
\usepackage{amssymb}

\usepackage{color}

{
{
{
\newcommand{\bea}{\begin{eqnarray}}
\newcommand{\eea}{\end{eqnarray}}

\newcommand{\nc}{\newcommand}
\nc{\renc}{\renewcommand}
\nc{\eqs}[2]{\mbox{Eqs.~(\ref{#1},\,\ref{#2})}}
\nc{\eq}[1]{\mbox{Eq.~(\ref{#1})}}
\nc{\figs}[2]{\mbox{Figs.~(\ref{#1},\,\ref{#2})}}
\nc{\fig}[1]{\mbox{Fig~.(\ref{#1})}}
\nc{\be}[1]{\begin{equation} \mbox{$\label{#1}$}}
\nc{\ee}{\vspace{0.1cm}\end{equation}}

\newcommand{\bean}{\begin{eqnarray*}}
\newcommand{\eean}{\end{eqnarray*}}

\def\GeV{{\rm \ GeV}}

\def\lae{\;^{<}_{\sim} \;} \def\gae{\; ^{>}_{\sim} \;}

\begin{document}

\title{Conventional and Unitarity-Conserving Peccei-Quinn Inflation Models and ACT} 

\author{John McDonald }
\email{j.mcdonald@lancaster.ac.uk}
\affiliation{Dept. of Physics,  
Lancaster University, Lancaster LA1 4YB, UK}

\begin{abstract} 

We compare conventional non-minimally coupled Peccei-Quinn (PQ) inflation with a version of the model in which unitarity conservation is imposed by additional Jordan frame interactions. Assuming instantaneous reheating, the unitarity-conserving model is within 1$\sigma$ agreement with the central value of the scalar spectral index reported by the ACT collaboration, whereas conventional PQ inflation is more than 2$\sigma$ below the ACT central value. In the case where dark matter is composed of axions and PQ symmetry is not restored after inflation, the axion isocurvature constraint of the unitarity-conserving model typically allows a much larger axion decay constant $f_{a}$ than the conventional model, with the conventional model upper bound being comparable only if the PQ scalar self-coupling is extremely small, $\lambda \lae 10^{-10}$. For  $\lambda = 0.1$, the axion isocurvature upper bounds are $f_{a} \leq 1.1 \times 10^{9} \GeV$ for conventional PQ inflation and $f_{a} \leq 6.4 \times 10^{13} \GeV$ for unitarity-conserving PQ inflation, with the latter bound being independent of $\lambda$. We also find a new isocurvature upper bound for conventional PQ inflation which is 650 times smaller than the existing bound. A modest reduction of the reheating temperature of the unitarity-conserving model from its maximum possible value will ensure that the PQ symmetry is not restored  after inflation, allowing values of $f_{a}$ up to $6.4 \times 10^{13} \GeV$. Thus only the unitarity-conserving PQ inflation model  allows $f_{a}$ to access values greater than the symmetry restoration cosmological upper bound     $\sim 10^{12} \GeV$ with naturally large values of the PQ scalar self-coupling.

\end{abstract} 
 
\maketitle

\section{Introduction}

Non-minimally coupled inflation models \cite{salopek} allow the inflaton to have large couplings comparable to those of the Standard Model (SM) and its particle physics extensions. This opens up the attractive possibility that inflation can be due to a conventional scalar field extension of the SM or even the SM Higgs boson itself \cite{bs}. However, there is a concern that such models may be fundamentally inconsistent as a result of unitarity violation \cite{uv1,uv2}. In the case of Higgs Inflation, the non-minimal coupling causes Higgs scattering via graviton exchange in the electroweak vacuum to violate perturbative unitarity at $E \sim M_{Pl}/\xi$, where $\xi$ is the non-minimal coupling. Whilst there is evidence to support the possibility that unitarity is conserved non-perturbatively and that Higgs Inflation is therefore a self-consistent theory \cite{sc1,sc2}, it is also possible that the theory will break down at energies greater than $M_{Pl}/\xi$ and so must be altered at higher energies to conserve unitarity. In this case it is not certain that inflation based on the SM Higgs potential will still be valid.

Recently we argued that unitarity conservation can be imposed on non-minimally coupled inflation models by including additional derivative interactions in the Jordan frame \cite{jmact}. In the Einstein frame the non-minimal kinetic term of the inflaton is replaced by a canonical kinetic term, eliminating unitarity violation. The inflaton potential is unchanged. The requires that the inflaton is a gauge singlet scalar, as quantum corrections due to gauge covariant derivatives in the Einstein frame would introduce unsuppressed quantum corrections to the inflaton potential due to gauge boson loops \cite{jmact,ucrl}. 

Support for unitarity-conserving non-minimally coupled inflation is provided by a new $\Lambda$CDM-based CMB determination of $n_s$ by the ACT collaboration\footnote{Other well-motivated inflation models have been proposed that are consistent with the ACT spectral index; see for example \cite{actm1}-\cite{na18} .} \cite{act}. Combining CMB data from Planck and ACT DR6 with DESI DR2 BAO data (the P-ACT-LB2 data set), the ACT collaboration finds $n_s = 0.9752 \pm 0.0030$. For the case of instantaneous reheating and with inflation self-coupling $\lambda = 0.1$, unitarity-conserving non-minimally coupled inflation predicts that $n_{s} = 0.9732$ and $r = 8.37 \times 10^{-6}$, with $n_{s}$ being well within 1$\sigma$ of the ACT collaboration central value \cite{jmact,ucrl}. 

In order to justify the additional Jordan frame interactions on the principle that unitarity must be conserved by a consistent quantum field theory, it is actually {\it necessary} that unitarity is violated by the non-minimal coupling of the scalar to the Ricci scalar. Since a non-minimally coupled real scalar field conserves unitarity \cite{rl0,rluv}, this suggests that the inflaton field is complex. 
A well-motivated candidate for a complex gauge singlet scalar extension of the SM is the Peccei-Quinn (PQ) scalar. PQ inflation via a conventional non-minimally coupled PQ scalar has previously been studied in \cite{fairb}. Here we will revisit this model and compare the predictions of the conventional PQ inflation model with those of the unitarity-conserving PQ inflation model.

\section{Conventional and Unitarity-Conserving PQ Inflation}

\subsection{Conventional PQ Inflation}

For conventional PQ inflation, the non-minimally coupled action in the Jordan frame is 
\be{e1} S_{J} = \int d^{4}x \sqrt{-g} \left( -\frac{M_{Pl}^{2} R}{2} - \xi \Phi^{\dagger} \Phi R + g^{\mu \nu} \partial_{\mu} \Phi^{\dagger} \partial_{\nu} \Phi - V(|\Phi|) \right) ~,\ee
where 
\be{e2} V(|\Phi|) = \lambda \left(|\Phi|^2 - \frac{f_{a}^{2}}{2}\right)^{2}   ~.\ee  
When transformed to the Einstein frame via the conformal transformation $\tilde{g}_{\mu \nu} = \Omega^{2} g_{\mu \nu}$, 
the action becomes 
\be{e3}  S_{E} = \int d^{4}x 
 \sqrt{-\tilde{g} } 
\left(
 - \frac{M_{Pl}^{2}}{2} \tilde{R} + 
\frac{1}{\Omega^2}\tilde{g}^{\mu \nu} \partial_{\mu} \Phi^{\dagger} \partial_{\nu} \Phi 
+ \frac{3 \xi^{2}}{\Omega^{4}  M_{pl}^{2} } 
\tilde{g}^{\mu \nu} \partial_{\mu}(\Phi^{\dagger} \Phi) \partial_{\nu}(\Phi^{\dagger} \Phi) -V_{E}(|\Phi|) \right)
~,\ee 
where 
\be{e4} \Omega^2 = 1 + \frac{2 \xi |\Phi|^{2}}{M_{Pl}^2}  ~\ee
and 
\be{e5} V_{E}(|\Phi|) = \frac{V(|\Phi|)}{\Omega^{4}}   ~.\ee
We will confirm that during inflation $|\Phi|^2 \gg f_{a}^2/2$, in which case $V(|\Phi|) \approx \lambda |\Phi|^4$.

\subsection{Unitarity Violation in Conventional PQ 
inflation} 

Let $\Phi = (\phi_{1} + i \phi_{2})/\sqrt{2}$ and assign the background inflaton as $\overline{\phi}_{1}$. Then when $\xi > 1$ the dominant energy scale of unitarity violation in the Einstein frame is $\Lambda \approx M_{Pl}/\xi$ for all background fields $\overline{\phi}_{1}$ \cite{rluv}. (We review the derivation of this result in Appendix A.)  
The theory must therefore be modified at an energy less than $\Lambda$ in order to be consistent with unitarity. Since the inflaton field is during inflation is larger than $M_{Pl}/\sqrt{\xi}$, this means that the inflaton potential is ill-defined during inflation if $\xi > 1$, which true unless the self-coupling $\lambda < 4.7 \times 10^{-10}$. This is because the effective theory would be expected to include inflaton potential terms suppressed by the scale of new physics, which would dominate the potential during inflation. In addition, quantum corrections to the potential are calculated at a renormalisation energy scale of the order of the inflaton field, which requires knowledge of the complete theory.

\subsection{Unitarity-Conserving PQ Inflation} 

The Jordan frame action of the inflaton sector is  
\be{e1}  S_{J} = \int d^{4} x \sqrt{-g} \left( -\frac{M_{P}^{2} R}{2} - \xi \Phi^{\dagger} \Phi R + \frac{2 \xi \Phi^{\dagger} \Phi}{M_{Pl}^{2}} g^{\mu \nu} \partial_{\mu} \Phi^{\dagger} \partial_{\nu} \Phi - \frac{3 \xi^{2}}{\Omega^{2} M_{Pl}^{2}} g^{\mu \nu} \partial_{\mu}\left(\Phi^{\dagger}\Phi\right) \partial_{\nu}\left(\Phi^{\dagger}\Phi\right)  + g^{\mu \nu} \partial_{\mu} \Phi^{\dagger} \partial^{\nu} \Phi   - V(|\Phi|)  \right) ~,\ee
where the third and fourth terms are the additional derivative interaction terms that ensure unitarity conservation \cite{jmact,ucrl}. 

When transformed to the Einstein frame via the conformal transformation $\tilde{g}_{\mu \nu} = \Omega^{2} g_{\mu \nu}$,
the action becomes   
\be{e12} S_{E} = \int d^{4} x 
\left(
 -\frac{M_{Pl}^{2} \tilde{R}}{2} 
+ g^{\mu \nu} \partial^{\mu} \Phi^{\dagger} \partial^{\nu} \Phi - V_{E}(|\Phi|)  \right) ~,\ee 
where 
\be{e13} V_{E} = \frac{V(|\Phi|)}{\Omega^{4}} = \frac{\lambda \left(|\Phi|^{2}  - \frac{f_{a}^{2}}{2}\right)^{2}}{\left(1 + \frac{2 \xi |\Phi|^{2}}{M_{Pl}^{2}} \right)^{2} }  ~.\ee

\section{Inflation predictions}

\subsection{Conventional PQ inflation} 

In general, the canonically normalised inflaton $\sigma$ is related to $\phi$ by 
\be{e15} \frac{d \sigma}{d \phi} = \frac{\left(\Omega^{2} + \frac{6 \xi^{2} \phi^{2}}{M_{Pl}^{2}} \right)^{1/2}}{\Omega^{2}} ~. ~\ee 
In the analytical results in this subsection we consider the case where $\xi > 1$. A general numerical analysis of the model for all $\xi$ is presented in subsection C. During inflation, when $\phi > M_{Pl}/\sqrt{\xi}$, \eq{e15} has the solution
\be{e16} \sigma = \frac{\sqrt{6} M_{Pl}}{2} \ln\left(\frac{\xi \phi^{2}}{M_{Pl}^{2}} \right) ~, ~\ee
where we have defined $\sigma = 0$ at $\phi = M_{Pl}/\sqrt{\xi}$. The potential is then 
\be{e17} V \approx \frac{\lambda M_{Pl}^{4}}{4 \xi^{2}} \left(1 - \frac{2 M_{Pl}^{2}}{\xi \phi^{2}} \right) = \frac{\lambda M_{Pl}^{4}}{4 \xi^{2}} \left(1 - 2 \exp\left(-\frac{2 \sigma}{\sqrt{6} M_{Pl}}\right) \right) ~.  ~\ee
For slow-roll inflation this gives the standard results for the number of e-foldings of inflation as a function of the inflaton field, the slow-roll parameters, the scalar spectral index and the tensor-to-scalar ratio,   
\be{e18} N = \frac{3}{4} \exp\left(\frac{2 \sigma}{\sqrt{6} M_{Pl}}\right) ~, ~\ee 

\be{e18a} \phi = \sqrt{\frac{4N}{3}} \frac{M_{Pl}}{\sqrt{\xi}}   ~, ~\ee 
\be{e19} \eta = - \frac{4}{3} \exp\left(-\frac{2 \sigma}{\sqrt{6} M_{Pl}} \right) \equiv -\frac{1}{N} \;\;\;,\;\;\; \epsilon = \frac{4}{3} \exp\left(-\frac{4 \sigma}{\sqrt{6} M_{Pl}} \right) \equiv \frac{3}{4 N^{2}} ~, ~\ee
\be{e21} n_{s} \approx 1 + 2 \eta = 1 - \frac{2}{N} \;\;\;,\;\; r = 16 \epsilon = \frac{12}{N^{2}} ~. ~\ee 
Thus with $N_{*} = 57$ at the pivot scale (we discuss the number of e-foldings at the pivot scale in Appendix B), we obtain $n_{s} = 0.9649$ 
and $ r = 3.69 \times 10^{-3} $.   
The power spectrum is 
\be{e23} P_{R} = \frac{V}{24 \pi^{2} \epsilon M_{Pl}^{4}} = \frac{\lambda N^{2}}{72 \pi^{2} \xi^{2}} ~. ~\ee  
Therefore 
\be{e23a} \xi =  \frac{\lambda^{1/2} N}{\sqrt{72} \pi P_{R}^{1/2}} ~. ~\ee
Using the Planck value at the pivot scale \cite{planck}, $P_{R} \equiv A_{s} = 2.1 \times 10^{-9}$, this gives
\be{e24} \xi = 4.51 \times 10^{4} \, \lambda^{1/2} \left( \frac{N_{*}}{57} \right)  ~.\ee 
Thus $\xi > 1$ unless $\lambda < 4.9 \times 10^{-10}$. 

Inflation in this model ends when $|\eta| = 1$, therefore
\be{e26} \phi_{end} = \sqrt{\frac{4}{3}} \frac{M_{Pl}}{\xi}   ~\ee
and
\be{e25} \sigma_{end} = \frac{\sqrt{6} M_{Pl}}{2} \ln \left( \frac{4}{3} \right) ~. ~\ee
We note that, for $\xi > 1$, the canonically normalised inflaton field $\sigma$ during inflation is generally super-Planckian.

\subsection{Unitarity-conserving PQ inflation} 

In this case the $\phi$ field is canonically normalised throughout and so the potential for the canonically normalised inflaton is 
\be{e27} V = \frac{\lambda \phi^{4}}{4 \left(1 + \frac{\xi \phi^{2}}{M_{Pl}^{2}} \right)^{2} } ~. \ee
The slow-roll results in this case are \cite{jmact}
\be{e28} N = \frac{\xi \phi^{4}}{16 M_{Pl}^{4}} ~,  ~\ee
\be{e28a} \eta = -\frac{12 M_{Pl}^{4}}{\xi \phi^{4}} \equiv -\frac{3}{4 N} \;\;\;,\;\; \epsilon = \frac{8 M_{Pl}^{6}}{\xi^{2} \phi^{6}} \equiv \frac{1}{8 \xi^{1/2} N^{3/2}} ~,  ~\ee 
\be{e29} n_{s} \approx 1 + 2 \eta = 1 -\frac{3}{2 N}  \;\;\;,\;\; r = \frac{2}{\sqrt{\xi} N^{3/2}} ~. ~\ee
The power spectrum is   
\be{e31} P_{R} = \frac{\lambda N^{3/2}}{12 \pi^2 \xi^{3/2}} ~\ee  
and so
\be{e32} \xi = \frac{ \lambda^{2/3} N}{\left(12 \pi^{2} P_{R}\right)^{2/3} } ~. ~\ee
With $P_{R} \equiv A_{s} = 2.1 \times 10^{-9}$ this gives
\be{e33} \xi =   1.40 \times 10^{6} \; \lambda^{2/3} \left( \frac{N_{*}}{56} \right) ~. ~\ee 
In this case $r$ is dependent upon $\lambda$,   
\be{e34} r = 3.89 \times 10^{-6} \; \lambda^{-1/3} ~. ~\ee
With $N_{*} = 56$, which is a good estimate for the case of instantaneous reheating for this model (Appendix B), and $\lambda = 0.1$, we find $n_{s} = 0.9732$ and $r = 8.37 \times 10^{-6}$.  

At end of inflation, corresponding to $|\eta| = 1$, the inflaton field is
\be{e35} \phi_{end} = \left(\frac{12}{\xi}\right)^{1/4} M_{Pl}  =  0.054 \left(\frac{56}{N_{*}}\right)^{1/4} \frac{M_{Pl}}{\lambda^{1/6}} ~. ~\ee
The canonically normalised inflaton field in this case is sub-Planckian throughout inflation if $0.054/\lambda^{1/6} < 1$, which is true if $\lambda > 2.5 \times 10^{-8}$. This is easily satisfied with dimensionally naturally large couplings, $\lambda \gae 10^{-3}$. This may be significant as quantum gravity in the Einstein frame becomes important at $E \sim M_{Pl}$. Depending on the details of the complete quantum gravity theory, the inflaton potential at field values greater than $M_{Pl}$ may be modified. Therefore unitarity-conserving PQ inflation may be preferred by consistency with quantum gravity.    

\subsection{General Analysis of Inflation} 

The above results are true if $\xi > 1$, in which case the non-minimal coupling is strong enough to dominate the dynamics of inflation. At sufficiently small $\xi$, however, the non-minimal coupling will become negligible in its effect and the results both models will tend towards $\phi^{4}$ chaotic inflation. 

We have integrated the slow-roll equation numerically for an arbitrary $\xi$. In this we have normalised the power spectrum to its observed value at the Planck pivot scale assuming instantaneous reheating by adjusting $\lambda$ for each value of $\xi$. The end of inflation is determined by either $|\eta| = 1$ or $\epsilon = 1$, depending on which occurs first. 

In Figure 1 we show $n_{s}$ as a function of $\xi$. We also show the P-ACT-LB2 central value and the 1$\sigma$ and 2$\sigma$ bounds. The conventional PQ inflation model (dotted curve) is generally more than 2$\sigma$ below the ACT central value, whereas the unitarity-conserving model (solid curve) is in very good agreement with the central value when $\xi > 1$. The spectral index deviates strongly from the non-minimally coupled inflation predictions only once $\xi \lae 0.01$, corresponding to $\lambda \lae 10^{-12}$.   

In Figure 2 we show $r$ as a function of $\xi$. We also show the present 2$\sigma$ upper bound on $r$, $r < 0.036$ \cite{rbound}, and a line at $r = 10^{-3}$ to indicate the magnitude that will be probed by next generation CMB polarisation experiments, notably LiteBIRD \cite{litebird}. For $\xi \gae 20$, corresponding to $\lambda \gae 10^{-8}$, the unitarity-conserving model (solid curve) predicts a value of $r \lae 10^{-3}$ and therefore primordial gravitational waves that will be unobservable in next generation CMB experiments. However, there is a window in $\xi$ from around $0.01$ to 20 for which the unitarity-conserving model $n_{s}$ is still within the 2$\sigma$ ACT bound whilst $r$ is between $10^{-3}$ and the present experimental upper bound on $r$. 

In Figure 3 we show $\lambda$ as a function of $\xi$, demonstrating that for a given $\lambda$ the unitarity-conserving model predicts a larger value of $\xi$. This implies that the energy density during inflation is lower in the unitarity-conserving model than in the conventional model for a given $\lambda$. Therefore the reheating temperature in the case of instantaneous reheating will be lower in the unitarity-conserving model for a given $\lambda$. 

Whilst it is possible for the unitarity-conserving model to be in agreement with ACT even if $\xi$ is as small as 0.01, corresponding to $\lambda \sim 10^{-12}$, such small inflaton self-couplings are not in the spirit of non-minimally coupled inflation, which is motivated by allowing inflation to be supported by a scalar field with couplings typical of those of conventional particle physics models, such as the Higgs boson with self-coupling  $\lambda \approx 0.1$.  In the following we will consider $\lambda \geq 10^{-3}$ to be dimensionally natural and focus on the results for this case.

\begin{figure}[h]
\begin{center}
\hspace*{-1.0cm}\includegraphics[trim = -3.5cm 0cm 0cm 0cm, clip = true, width=0.55\textwidth, angle = -90]{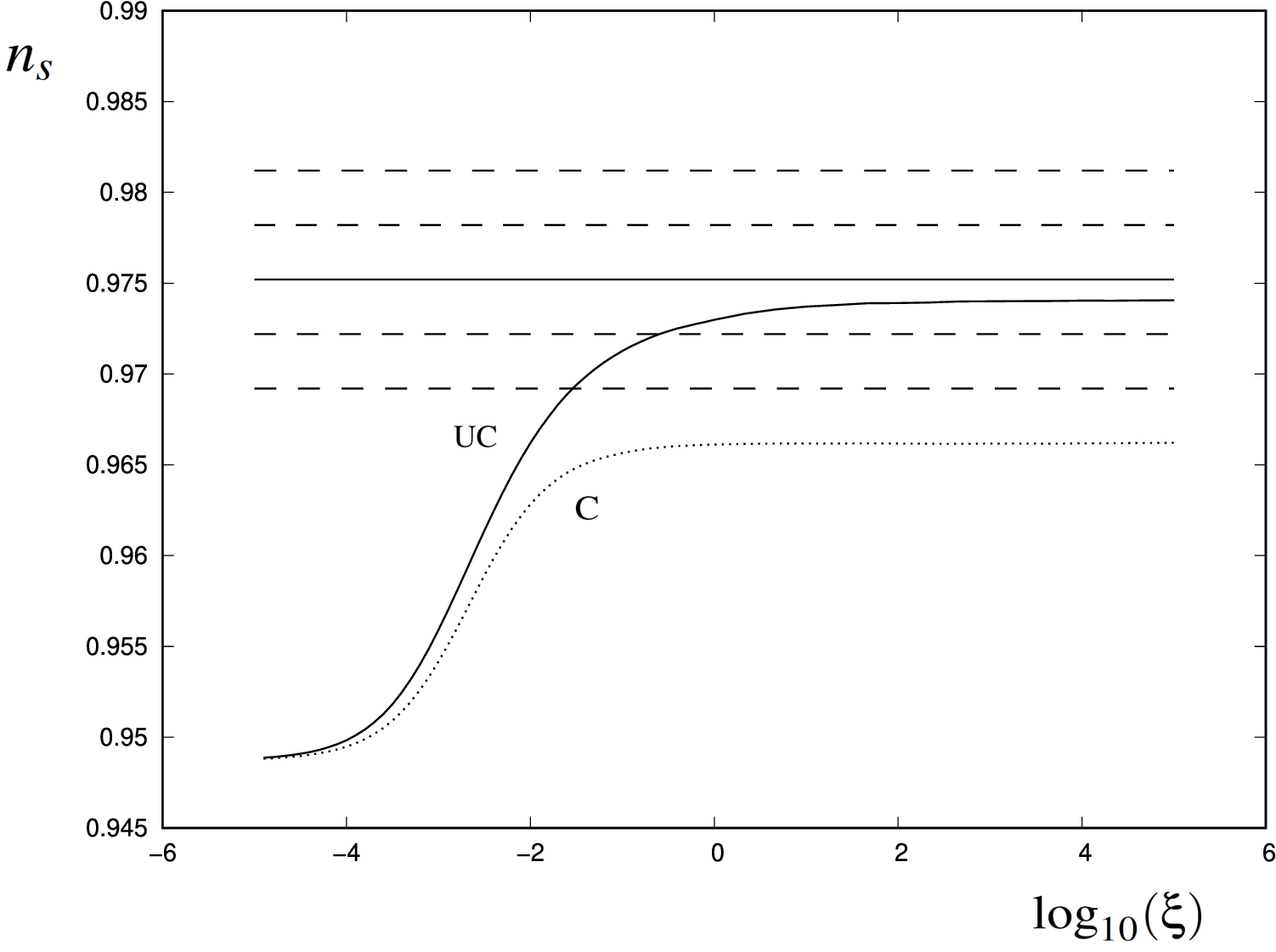}
\caption{Scalar spectral index $n_{s}$ as a function of the non-minimal coupling $\xi$ for the conventional PQ inflation model (dotted curve, C) and for the unitarity-conserving model (solid curve, UC). The central value and 1$\sigma$ and 2$\sigma$ bounds from the ACT collaboration P-ACT-LB2 analysis are also shown. }
\label{fig1}
\end{center}
\end{figure}

\begin{figure}[h]
\begin{center}
\hspace*{-1.0cm}\includegraphics[trim = -3.5cm 0cm 0cm 0cm, clip = true, width=0.55\textwidth, angle = -90]{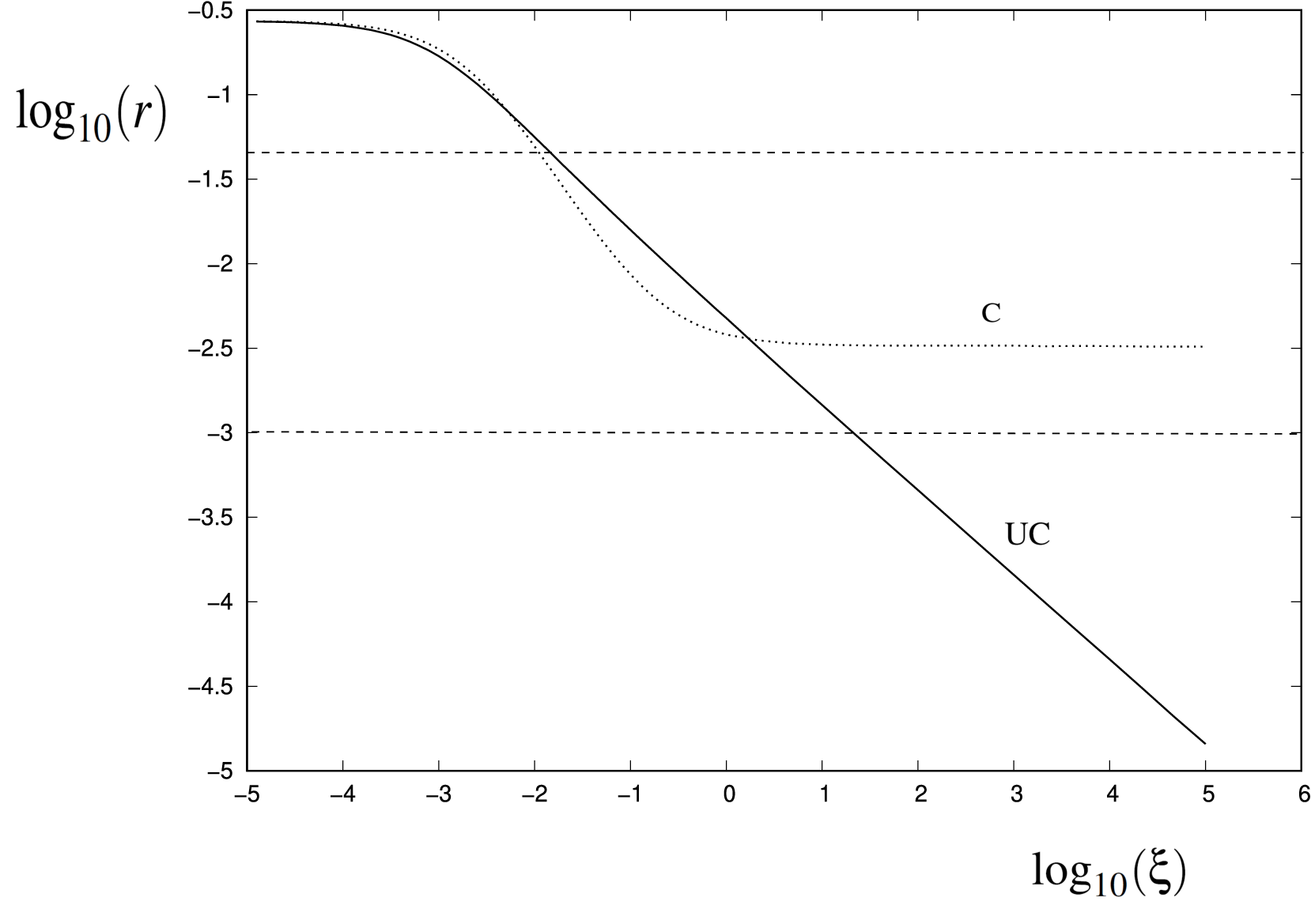}
\caption{Tensor-to-scalar ratio $r$ as a function of the non-minimal coupling $\xi$ for the conventional PQ inflation model (dotted curve, C) and the unitarity-conserving model (solid curve, UC). Dashed lines show the present observational upper bound, $r < 0.036$, and the magnitude that will be probed by next generation CMB experiments, $r = 10^{-3}$.} 
\label{fig2}
\end{center}
\end{figure}

\begin{figure}[h]
\begin{center}
\hspace*{-1.0cm}\includegraphics[trim = -3.5cm 0cm 0cm 0cm, clip = true, width=0.55\textwidth, angle = -90]{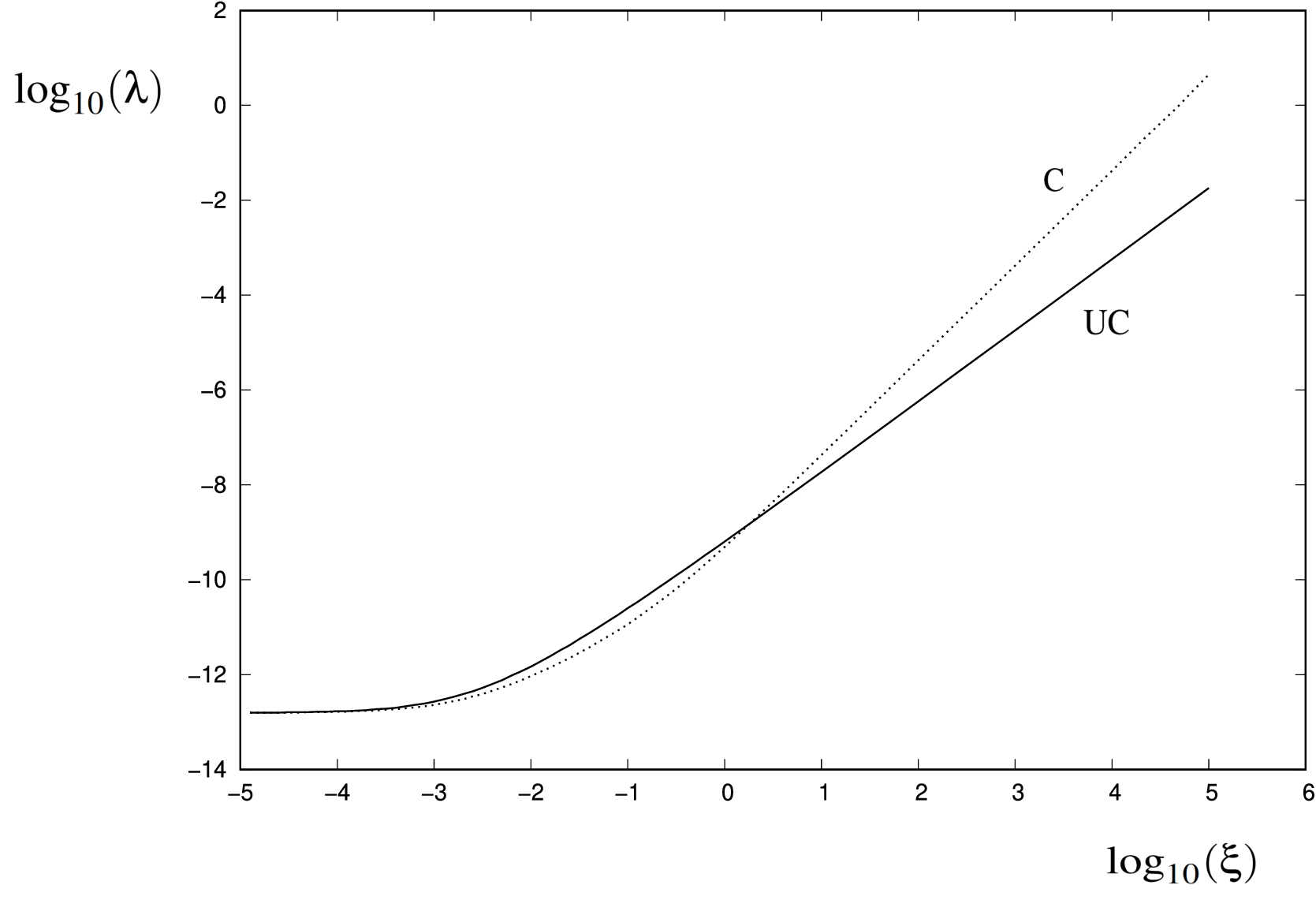}
\caption{Inflaton self-coupling $\lambda$ as a function of the non-minimal coupling $\xi$ for the conventional PQ inflation model (dotted curve, C) and the unitarity-conserving model (solid curve, UC).} 
\label{fig3}
\end{center}
\end{figure}

\section{Axion Isocurvature Constraints}  

\subsection{Conventional PQ inflation} 

Axion isocurvature perturbations in conventional PQ inflation were discussed in \cite{fairb}. In this section this we revisit this calculation. We find a much smaller upper bound on $f_{a}$ than that found in \cite{fairb}, for reasons that we will explain. 

We assume that the PQ symmetry is not restored after reheating, so that the angular component of $\Phi$ and its quantum fluctuations are not erased. We define the inflaton field radial direction as $\phi_{1}$. Fluctuations in the angular direction are then $\delta \phi_{2}$.
The PQ field kinetic term is 
\be{e37} \frac{1}{\Omega^{2}} \partial_{\mu} \Phi^{\dagger} \partial^{\mu} \Phi ~\ee 
and so the $\phi_{2}$ kinetic term is 
\be{e38} \frac{1}{2 \Omega^{2}} \partial_{\mu} \phi_{2} \partial^{\mu} \phi_{2} ~. ~\ee 
In the inflaton background $\overline{\phi}_{1}$ with $\xi \overline{\phi}_{1}^{2}/M_{Pl}^{2} \gg 1$, which applies when $\xi \gae 1$, the $\phi_{2}$ kinetic term becomes 
\be{e39} \frac{M_{Pl}^{2}}{2 \xi \overline{\phi}_{1}^{2} }  \partial_{\mu} \phi_{2} \partial^{\mu} \phi_{2} ~. ~\ee
We define a canonically normalised field $\chi_{2}$ as 
\be{e40} \chi_{2} = \frac{M_{Pl} \phi_{2}}{\sqrt{\xi} \overline{\phi}_{1}}  ~. ~\ee 
The angular field fluctuation is $\delta \theta = \delta \phi_{2}/\overline{\phi}_{1}$. Thus 
\be{e41} \delta \theta = \frac{\sqrt{\xi}\delta \chi_{2}}{M_{Pl}}  ~. ~\ee 
Therefore the power spectrum of $\theta$ fluctuations is 
\be{e42} P_{\delta \theta} = \frac{\xi}{M_{Pl}^{2}} P_{\delta \chi_{2}} ~. ~\ee 
Since $\chi_{2}$ is a canonically normalised massless scalar, its quantum fluctuation during inflation will take the usual form, with power spectrum
\be{e43} P_{\delta \chi_{2}} = \frac{H^{2}}{4 \pi^{2}} ~. ~\ee 
Therefore
\be{e44} P_{\delta \theta} = \frac{\xi}{M_{Pl}^{2}} \frac{H^{2}}{4 \pi^{2}} ~. ~\ee 
The axion isocurvature fluctuation due to $\delta \theta$ is 
\be{e45} I \equiv \frac{\delta \rho}{\rho} = \frac{2 \delta \theta}{\theta} ~, ~\ee
where $\rho_{a}$ is the axion density due to misalignment by $\theta$, assumed to be small compared to $\pi$,  
\be{e45a} \rho_{a} = \frac{1}{2} m_{a}^{2} f_{a}^{2} \theta^{2} ~ \ee
and $m_{a}$ is the axion mass. 
Therefore the power spectrum of $I$ is 
\be{e46} P_{I} = \frac{4 P_{\delta \theta}}{\theta^{2}}  = \frac{\xi H^{2}}{\pi^{2} \theta^{2} M_{Pl}^{2}} ~. ~\ee 
We assume that CDM is due to axions. The CDM isocurvature perturbation is parameterised by $\beta_{iso}$, defined as 
\be{e50a} \beta_{iso} = \frac{P_{I}}{P_{R} + P_{I}}  ~. ~\ee
Since $P_{R} \gg P_{I}$ is necessary to be consistent with observation,  we have, to a good approximation, 
\be{e46a}  \beta_{iso} = \frac{P_{I}}{P_{R}} = 
 \frac{\xi H^{2}}{\pi^{2} \theta^{2} M_{Pl}^{2} P_{R}} ~. ~\ee 
During inflation, and assuming that $\xi \gae 1$, $V_{E} = \lambda M_{Pl}^{4}/(4 \xi^{2})$ to a good approximation, therefore
\be{e47} H^{2} = \frac{\lambda M_{Pl}^{2}}{12 \xi^{2}}  ~.\ee 
Thus
\be{e48} \beta_{iso} = \frac{\lambda}{12 \pi^{2} \theta^{2} \xi P_{R}} ~. ~\ee 
Using the relation between $\lambda$ and $\xi$, \eq{e23a}, 
we obtain 
 \be{e51} \beta_{iso} = \frac{\lambda^{1/2}}{\sqrt{2} \pi \theta^{2} N P_{R}^{1/2}} ~. ~\ee

To compare with the results of \cite{fairb}, we use the same relation between the value of $\theta$ and the axion dark matter density as used in \cite{fairb} 
\be{e51} \theta^{2} = 6 \times 10^{-6} \left( \frac{\Omega_{a} h^{2}}{0.1199} \right) \left( \frac{10^{16} \GeV}{f_{a}}\right)^{7/6}  ~. ~\ee 
Therefore, at the pivot scale,   
\be{e52} \beta_{iso} = 1.5 \times 10^{7} \, \lambda^{1/2} 
\left( \frac{0.1199}{\Omega_{a} h^{2}} \right) \left( \frac{f_{a}}{10^{16} \GeV}\right)^{7/6}\left(\frac{57}{N_{*}}\right) ~. ~\ee 
The observational 2$\sigma$ upper bound is $\beta_{iso} < 0.038$ \cite{planck}. Therefore
\be{e53} f_{a} < \frac{4.3 \times 10^{8}}{\lambda^{3/7}} 
\left( \frac{\Omega_{a} h^{2}}{0.1199} \right)^{6/7} \left(\frac{N_{*}}{57}\right)^{6/7} \, \GeV ~. ~\ee 
We can also express this as an upper bound on $\lambda$ for a given $f_{a}$ 
\be{e54} \lambda < \left(\frac{4.3 \times 10^8 \GeV}{f_{a}}\right)^{7/3} \left(\frac{\Omega_{a} h^{2}}{0.1199} 
\right)^{2} \left(\frac{N_{*}}{57}\right)^{2} ~. ~\ee 

The present lower bound on the axion decay constant from astrophysics (stellar cooling, SN1987a) is $f_{a} \gae 10^{9}\, \GeV$. In addition, if PQ symmetry is restored after inflation there is a cosmological upper bound from the axion dark matter density due to coherent oscillations and string and domain wall decay,  $f_{a} \lae 10^{12} \GeV$.

From \eq{e54} we find that $f_{a} = 10^{9} \GeV \Rightarrow \lambda = 0.13$ in the conventional PQ inflation model. Therefore if the value of $\lambda$ is of the order of 0.1, typical of particle physics models, then the axion decay constant will be very close to the present astrophysical bound and so astrophysical axions may be close to observable. 

In addition, in order for the conventional PQ inflation model to have a value larger than the cosmological upper bound, a very small value for $\lambda$ would be necessary, since $f_{a} > 10^{12} \GeV$ requires that $\lambda < 1.3 \times 10^{-8}$.  Dimensionally naturally large couplings, $\lambda \geq 10^{-3}$, require that $f_{a} \leq 8.1 \times 10^{9} \GeV$.

\begin{figure}[h]
\begin{center}
\hspace*{-1.0cm}\includegraphics[trim = -3.5cm 0cm 0cm 0cm, clip = true, width=0.55\textwidth, angle = -90]{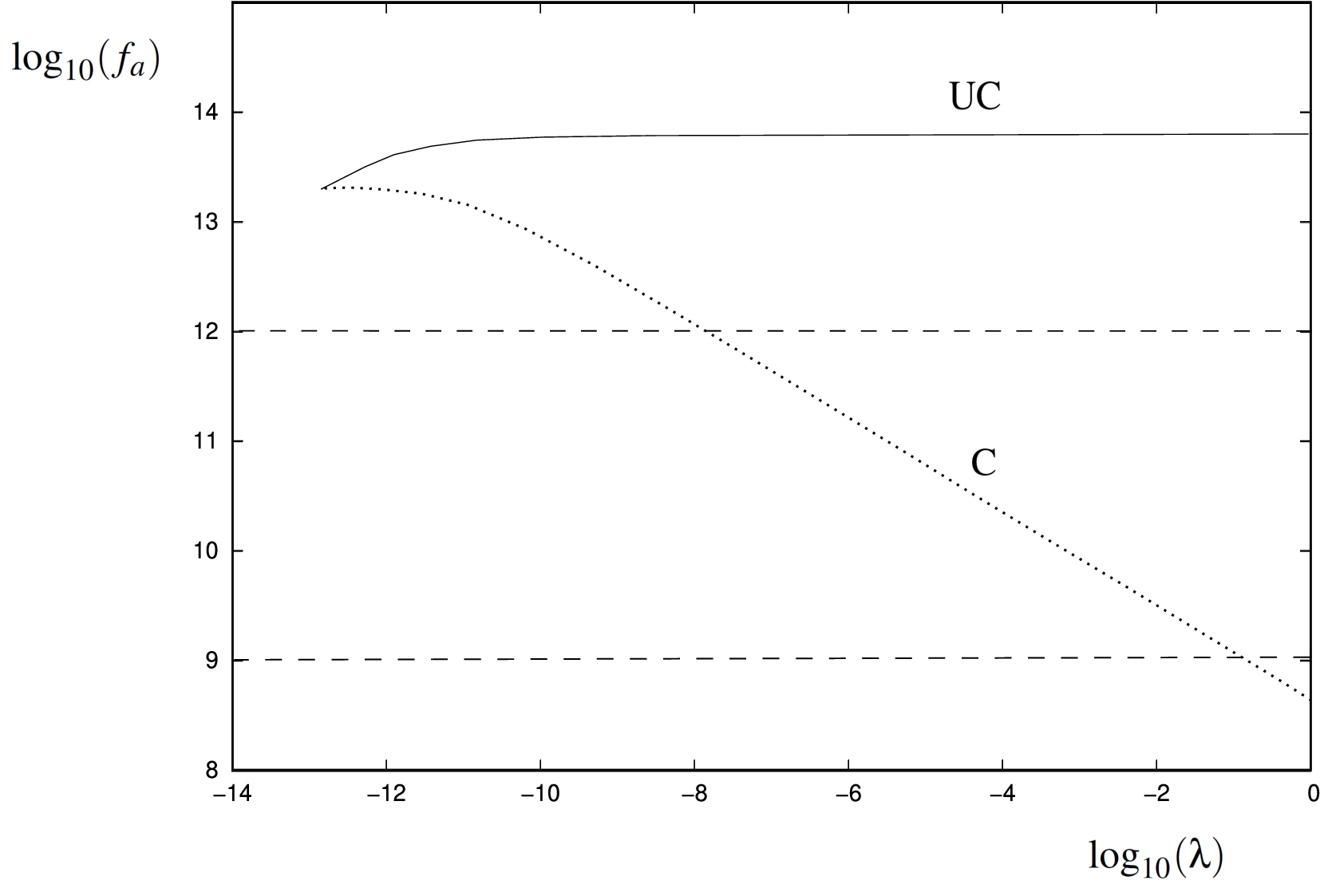}
\caption{Axion isocurvature upper bound on $f_{a}$ in GeV for conventional PQ inflation (dotted line, C) and for unitarity-conserving PQ inflation (solid line, UC). The dashed line at $f_{a} = 10^{9} \GeV$ indicates the astrophysical lower bound and the dashed line at $f_{a} = 10^{12} \GeV$ indicates the cosmological upper bound for the case where PQ symmetry is restored after inflation. At small $\lambda$ both upper bounds converge to that of $\phi^4$ inflation. } 
\label{fig4}
\end{center}
\end{figure}

We next compare our result with that obtained in \cite{fairb}. Their result is equivalent to a value of $\beta_{iso}$, which we call $\beta_{iso,\,old}$, given by (from Equation 18 of \cite{fairb})  
\be{e55}  \beta_{iso,\, old} = \frac{M_{Pl}^{2} \epsilon} {\pi s_{*}^{2} \theta^{2}}\equiv \frac{H^{2}}{8 \pi^{3} \overline{\phi}_{1}^{2} \theta^{2} P_{R}} ~,\ee
where their $s_{*}$ is equivalent to our $\overline{\phi}_{1}$. 
Comparing with our $\beta_{iso}$, \eq{e46a}, we find that our value is enhanced by a factor $\kappa_{e}$, where
\be{e56} \kappa_{e} =  \frac{\beta_{iso}}{\beta_{iso,\,old}} = \frac{ 8 \pi \xi \overline{\phi}_{1}^{2} }{M_{Pl}^{2}} = \frac{32 \pi N}{3} ~. ~\ee  
In the final expression we used the relation \eq{e18a} between $\overline{\phi}_{1}$ and $N$. So with $N = 57$ we find that our isocurvature parameter is larger by a factor $\kappa_{e} = 1910$. 

Since both $\beta_{iso}$ and $\beta_{iso,\,old}$ are proportional to $1/\theta^{2}$ and so, from \eq{e51}, proportional to $f_{a}^{7/6}$, it follows that the axion isocurvature upper bound on $f_{a}$ in \cite{fairb} will be larger than ours by a factor $\kappa_{e}^{6/7} \approx 650$. 

We have identified two reasons for the difference between our result and that of \cite{fairb}. The first is that the analysis of \cite{fairb} treats the angular field (equivalent to our $\delta \phi_{2}$) as a canonically normalised field throughout and overlooks the effect of $\Omega$ on the kinetic term during inflation. The second is that there is a missing factor of 8$\pi$ in the expression for the isocurvature perturbation fraction in \cite{fairb}, which assumes an expression for the isocurvature perturbation fraction given in \cite{gba}. In the expression in \cite{gba}, the Planck mass is the true Planck mass, $1/G^{1/2}$, but this is interpreted as reduced Planck mass in \cite{fairb}.

\subsection{Unitarity-conserving PQ inflation}  

In this case the axion field is canonically normalised throughout, therefore
\be{e57} \beta_{iso} = \frac{H^{2}}{\pi^{2} \overline{\phi}_{1}^{2} \theta^{2} P_{R}} ~.  ~\ee 
With $H^{2} = \lambda M_{Pl}^{2}/12 \xi^{2}$, which applies when $\xi \gae 1$, this gives
\be{e58} \beta_{iso} = \frac{\lambda M_{Pl}^{2}}{12 \xi^{2} \pi^{2} \theta^{2} \overline{\phi}_{1}^{2} P_{R} } ~.  ~\ee 
Using the relation \eq{e32} between $\xi$ and $\lambda$ and the expression \eq{e28} for $\phi_{1}$ as a function of $N$, we find the simple result 
\be{e59} \beta_{iso} = \frac{1}{4 \theta^{2} N^{2}}  ~. ~\ee 
Thus in this case the isocurvature perturbation is independent of $\lambda$ and $\xi$. Using the expression \eq{e51} for $\theta^2$ gives
\be{e60} \beta_{iso} = 13.6 \left(\frac{56}{N_{*}}\right)^{2} \left( \frac{0.1199}{\Omega_{a} h^{2}} \right) \left(\frac{f_{a}}{10^{16} \GeV}\right)^{7/6} ~. ~\ee
Therefore $\beta_{iso} < 0.038$ implies that 
\be{e61} f_{a} < 6.4 \times 10^{13} \left(\frac{N_{*}}{56}\right)^{12/7} \left( \frac{\Omega_{a} h^{2}}{0.1199}\right)^{6/7} \GeV ~. ~\ee 

We show the axion isocurvature upper bound for the conventional and unitarity-conserving models as a function of $\lambda$ in Figure 4. In this figure we have numerically generalised the analytical upper bounds for $\xi \gae 1$ to any value of $\xi$ by using the complete expressions for $\Omega$ and $V$ without approximation. The upper bounds of both models converge to the $\phi^4$ inflation upper bound at very small $\lambda$.  The upper bound is much larger for unitarity-conserving PQ inflation unless  $\lambda$ is very small compared to 1. In addition, the upper bound for unitarity-conserving PQ inflation is generally greater than 
the symmetry restoration cosmological upper bound, whereas for conventional PQ inflation this would require that $\lambda \lae 10^{-8}$. Therefore, for dimensionally natural couplings, $\lambda \geq 10^{-3}$, only the unitarity-conserving model can access values of $f_{a}$ greater than the cosmological upper bound. 

The upper bound on $f_{a}$, $6.4 \times 10^{13} \GeV$, is generally small compared to the PQ field $\phi$ at the end of inflation. The smallest PQ field at the end of inflation is that of the conventional PQ inflation model, for which $\phi_{end} = (4/3)^{1/2} M_{Pl}/\sqrt{\xi} \gae 1.3 \times 10^{16} \GeV$, where we have used the value of $\xi$ from \eq{e24} with $\lambda = 1$ as the largest value of $\xi$. Therefore the assumption made in our analysis of inflation, that $\phi^{2} \gg f_{a}^{2}$, is easily satisfied.

\section{PQ Symmetry Non-Restoration}

\subsection{Conventional PQ Inflation} 

The highest reheating temperature corresponds to instantaneous reheating at the end of inflation. In the following we assume that $\xi \gae 1$. With $\rho_{rad} = V(\phi_{end}) = \lambda M_{Pl}^{4}/(4 \xi^{2})$, we obtain
\be{rh1} T_{R,\;max} =  \frac{ k_{T_{R}} \lambda^{1/4} M_{Pl}}{\sqrt{\xi}} ~, ~\ee
where $k_{T_{R}} = (15/(2 \pi^{2} g(T_{R}) ))^{1/4} = 0.29$ for $g(T_{R}) = 106.75$.

In the case of conventional PQ inflation, substituting the relation between $\xi$ and $\lambda$ , \eq{e24}, we find 
\be{rh2} T_{R,\,max} = 3.2 \times 10^{15} \left(\frac{57}{N_{*}}\right)^{1/2} \, \GeV ~. ~\ee
The condition for symmetry restoration after inflation is that 
$T_{RH} > \sqrt{2} f_{a}$ \cite{fairb}. This ensures that the temperature dependent $\phi^2$ term in the finite-temperature effective potential due to the self-coupling, $\lambda T^2 \phi^{2}/4$, is larger in magnitude than the symmetry breaking $\phi^2$ term, $-\lambda f_{a}^{2}\phi^{2}/2$.     
We assume that the actual reheating temperature is less $T_{R,\,max}$ by a factor $\gamma$, $T_{R} = \gamma T_{R,\,max}$. Then the condition on $\gamma$ for symmetry not to be restored in the case of conventional PQ inflation is 
\be{rh3} \gamma < \frac{f_{a}}{2.3 \times 10^{15} \GeV}\left(\frac{N_{*}}{57}\right)^{1/2} ~. ~\ee
Therefore, since $f_{a} \leq 8.1 \times 10^{9} \GeV$ if $\lambda \geq 10^{-3}$, a suppression of the reheating temperature relative to its maximum value by $\gamma < 3.5 \times 10^{-6}$ would be necessary to prevent symmetry restoration if 
$\lambda \geq 10^{-3}$.  

\subsection{Unitarity-Conserving PQ Inflation}

In the case of unitarity-conserving PQ inflation, using the relation between $\xi$ and $\lambda$, \eq{e33}, we obtain 
\be{rh3} T_{R,\,max} = \frac{5.8 \times 10^{14} \GeV}{\lambda^{1/12}} \left( \frac{56}{N_{*}}\right)^{1/2} ~. ~\ee  
Therefore the condition for symmetry not to be restored in this case is 
\be{rh4} \gamma < \frac{\lambda^{1/12} f_{a}}{4.1 \times 10^{14} \GeV}\left(\frac{N_{*}}{56}\right)^{1/2} ~. ~\ee
Thus, with only a moderate suppression of the reheating temperature relative to its maximum value, it is possible to have unrestored PQ symmetry and $f_{a}$ larger than the cosmological upper bound. For example, with $\lambda = 0.1$, $f_{a} > 10^{13} \GeV$ is consistent with no symmetry restoration if $\gamma < 0.02$, whilst the isocurvature upper bound, $f_{a} = 6.4 \times 10^{13} \GeV$, is consistent if $\gamma < 0.13$. Such a suppression of the reheating temperature is always possible if the couplings of the PQ scalar to the heavy quarks of the axion sector and to the SM Higgs are sufficiently small. 

\section{Conclusions} 

We have revisited the conventional non-minimally coupled PQ inflation model and compared its predictions with those of a unitarity-conserving version of the model. Our main results are:
\newline (i) The scalar spectral index of the conventional PQ inflation model is more that 2$\sigma$ below the ACT central value of $n_{s}$, whereas the unitarity-conserving model is well within 1$\sigma$ agreement with ACT, assuming instantaneous reheating.
\newline (ii) When the PQ self-coupling is dimensionally natural, $\lambda \geq 10^{-3}$, the tensor-to-scalar ratio in the conventional model is $r \approx 4 \times 10^{-3}$,  whereas it is less than $4 \times 10^{-5}$ in the unitarity-conserving model. The unitarity-conserving model produces $r > 10^{-3}$ only when $\lambda \lae 10^{-8}$. Therefore $r$ is potentially observable in next-generation CMB polarisation experiments (which will probe down to $r \sim 10^{-3}$) only in the conventional PQ inflation model,  unless the PQ scalar self-coupling is extremely small. 
\newline  (iii) The axion isocurvature upper bound on $f_{a}$ in the conventional model is less than $10^{10} \GeV$ for $\lambda \geq 10^{-3}$ and becomes greater that $10^{12} \GeV$ only if $\lambda \lae 10^{-8}$, whereas the upper bound for the unitarity-conserving model is $6.4 \times 10^{13} \GeV$ and is independent of $\lambda$ except for a slight reduction at $\lambda \lae 10^{-11}$. Therefore only the unitarity-conserving model can have an axion decay constant larger than the cosmological upper bound, $\sim 10^{12} \GeV$, unless the PQ scalar coupling is extremely small. 
\newline (iv) The canonically normalised Einstein frame inflaton in the conventional PQ inflation model is generally  super-Planckian during inflation, whereas in the unitarity-conserving PQ inflation model it is sub-Planckian if $\lambda \gae 10^{-8}$. Therefore the unitarity-conserving model may be preferred by quantum gravity. 

We also found a new result for the axion isocurvature upper bound on $f_{a}$ in the conventional PQ inflation model, which is around 650 times smaller than the existing upper bound.

   \renewcommand{\theequation}{A-\arabic{equation}}
 \setcounter{equation}{0}  

\section*{Appendix A: Unitarity Violation in Conventional PQ Inflation} 

To explain the energy scale of unitarity violation, we follow the analysis presented in \cite{jmact} and \cite{rluv}. In the case where $\overline{\phi}_{1} > M_{Pl}/(\sqrt{6} \xi)$ and $\xi > 1$, the term that leads to the dominant unitarity-violating interaction is the third term in \eq{e3}, 
\be{aa1} \frac{3 \xi}{\Omega^{4} M_{Pl}^{2}} \partial_{\mu}(\Phi^{\dagger} \Phi)\partial^{\mu}(\Phi^{\dagger} \Phi) ~. ~\ee
We expand around the background inflaton field as 
\be{aa2} \Phi = \frac{1}{\sqrt{2}} \left( \overline{\phi}_{1} + \delta \phi_{1} + i \delta \phi_{2} \right) ~\ee 
and Taylor expand \eq{aa1}
\be{aa3} \frac{3 \xi^{2}}{\Omega^{4} M_{Pl}^{2}} \left[ \left(\overline{\phi}_{1}^{2} + 2 \overline{\phi}_{1} \delta \phi_{1} + \delta \phi_{1}^{2} \right) \partial_{\mu} \delta \phi_{1} \partial^{\mu} \delta \phi_{1}
+ 2 \left(\overline{\phi}_{1} + \delta \phi_{1}\right) \delta \phi_{2} \partial_{\mu} \delta \phi_{1} \partial^{\mu} \delta \phi_{2}
+ 4 \delta \phi_{2}^{2} \partial_{\mu} \delta \phi_{2} \partial^{\mu} \delta \phi_{2} \right] ~.  ~\ee 
This modifies the kinetic terms of $\delta \phi_{1}$ and $\delta \phi_{2}$ and introduces an interaction between $\delta \phi_{1}$ and $\delta \phi_{2}$. The kinetic terms and the dominant interaction term are 
\be{aa4} \frac{1}{2 \Omega^{2}} \left(1 + \frac{6 \xi^{2} \overline{\phi}_{1}^{2}}{M_{Pl}^{2} \Omega^{2}} \right) \partial_{\mu} \delta \phi_{1} \partial^{\mu} \delta \phi_{1} 
+ \frac{1}{2 \Omega^{2}}  \partial_{\mu} \delta \phi_{2} \partial^{\mu} \delta \phi_{2}  + \frac{6 \xi^{2}}{\Omega^{4} M_{Pl}^{2}} \overline{\phi}_{1} \delta \phi_{2} \partial_{\mu} \delta \phi_{1} \partial^{\mu} \delta \phi_{2} ~. ~\ee 
We define canonically normalised fields in the background $\overline{\phi}_{1}$ by 
\be{aa5} \delta \varphi_{1} = \frac{\left(1 + \frac{6 \xi^{2} \overline{\phi}_{1}^{2}}{M_{Pl}^{2} \Omega^{2}} \right)^{1/2}\delta \phi_{1} }{ \Omega}   ~\ee 
and 
\be{aa6} \delta \varphi_{2} = \frac{\delta \phi_{2}}{\Omega}  ~, \ee 
where 
\be{aa7} \Omega^{2} = \left(1 + \frac{\xi \overline{\phi}_{1}^{2}}{M_{Pl}^{2}} \right) ~. ~\ee 
The interaction term then becomes 
\be{aa8} \frac{1}{\Lambda} \delta \varphi_{2} \partial_{\mu} \delta \varphi_{1} \partial^{\mu} \delta \varphi_{2}  ~, ~\ee 
where
\be{aa9} \Lambda = \frac{ \Omega M_{Pl}^{2}}{6 \xi^{2} \overline{\phi}_{1}} \left(1 + \frac{ 6 \xi^{2} \overline{\phi}_{1}^{2} }{M_{Pl}^{2} \Omega^{2}} \right)^{1/2} ~. ~\ee
This interaction allows $\delta \phi_{1} \delta \phi_{2} \rightarrow \delta \phi_{1} \delta \phi_{2}$ via $\delta \phi_{2}$ exchange, which will  violate unitarity at $E \gae \Lambda$. We assume $\xi > 1$ in the following. There are then two possibilities:

\noindent (i) $\xi \overline{\phi}_{1}^{2}/M_{Pl}^{2} > 1$ and $6 \xi^{2} \overline{\phi}_{1}^{2}/(M_{Pl}^{2} \Omega^{2}) > 1$. In this case we find that $\Lambda \approx M_{Pl}/(\sqrt{6} \xi)$.

\noindent (ii) $\xi \overline{\phi}_{1}^{2}/M_{Pl}^{2} < 1$ and $6 \xi^{2} \overline{\phi}_{1}^{2}/(M_{Pl}^{2} \Omega^{2}) > 1$. In this case, with $\Omega \approx 1$, we find again that $\Lambda \approx M_{Pl}/(\sqrt{6} \xi)$.

\noindent Once $6 \xi^{2} \overline{\phi}_{1}^{2}/(M_{Pl}^{2} \Omega^{2}) < 1$, the $\overline{\phi}_{1}$-dependent interaction in \eq{aa4} is no longer the dominant unitarity violating interaction. The dominant interaction is then the $\overline{\phi}_{1}$-independent interaction 
\be{e10} \frac{6 \xi^{2}}{M_{Pl}^{2}} \phi_{1} \phi_{2} \partial_{\mu}  \phi_{1}\partial^{\mu} \phi_{2} ~. ~\ee
This again violates unitarity at the energy scale $\Lambda = M_{Pl}/\sqrt{6} \xi$. Thus if $\xi > 1$, the energy scale of unitarity violation is independent of the background field and is given by $\Lambda = M_{Pl}/\sqrt{6} \xi$. 

\noindent If $\xi < 1$, then the dominant unitarity-violating interaction comes from expanding the kinetic terms, in which case $\Lambda \approx M_{Pl}/\sqrt{\xi}$ for all background fields \cite{jmpal}.

   \renewcommand{\theequation}{B-\arabic{equation}}
 \setcounter{equation}{0}  

\section*{Appendix B: $N_{*}$ for instantaneous reheating}

The number of e-foldings corresponding to the Planck pivot scale is determined from the time of horizon exit of the pivot scale wavenumber, corresponding to 
\be{bb1}  \left(\frac{a_{0}}{a}\right) \,k_{*} = 2 \pi H  ~,\ee
where $a$ is the scale factor and '0' denotes present value.
During inflation, and assuming that $\xi \gae 1$, for both the conventional and unitarity-conserving models $\rho \approx V_{0} = \lambda M_{Pl}^{4}/(4 \xi^{2})$. 
Therefore 
\be{bb2} H = \left(\frac{V_{0}}{3 M_{Pl}^{2}} \right)^{1/2} = \left(\frac{\lambda}{12}\right)^{1/2} \frac{M_{Pl}}{\xi}   ~.\ee
$V_{0}$ is assumed to convert to radiation at the end of inflation, therefore the reheating temperature is
\be{bb3} T_{R} = \left(\frac{30 V_{0}}{\pi^{2} g(T_{R})}\right)^{1/4}  ~,\ee
where $T_{0}$ is the present CMB temperature and $g(T)$ is the effective number of relativistic degrees of freedom.  
Thus the LHS of \eq{bb1} is 
\be{bb4} \left(\frac{a_{0}}{a}\right) \, k_{*} = \left(\frac{a_{0}}{a_R}\right) \left(\frac{a_{R}}{a}\right) k_{*} = \left(\frac{g(T_{R})}{g(T_{0})} \right)^{1/3} \left(\frac{T_{R}}{T_{0}}\right)\, e^{N_{*}}\, k_{*}  ~.\ee
From \eq{bb1}, \eq{bb2} and \eq{bb3} we obtain 
\be{bb5} e^{N_{*}} = \frac{2 \pi T_{0}}{\xi^{1/2} k_{*}} \left(\frac{g(T_{0})}{g(T_{R})^{1/4}} \right)^{1/3} \left(\frac{\pi^{2}}{90} \right)^{1/4} \left(\frac{\lambda}{12}\right)^{1/4}   ~.\ee

For the case of conventional PQ inflation we use \eq{e23a} to eliminate $\xi$ in terms of $\lambda$. 
\eq{bb5} then gives 
\be{bb7} e^{N_{*}} = \frac{c_{1} T_{0} P_{R}^{1/4}}{k_{*}}  \left(\frac{g(T_{0})}{g(T_{R})^{1/4}} \right)^{1/3} \frac{1}{N_{*}^{1/2}}   ~\ee
where 
\be{bb8} c_{1} = (4 \sqrt{6} \pi^3)^{1/2} \left(\frac{\pi^{2}}{90} \right)^{1/4} = 10.03  ~.\ee
Using the Planck pivot scale, $k_{*} = 0.05 \,{\rm Mpc}^{-1} \equiv 3.20 \times 10^{-40} \GeV$, $g(T_{0}) = 3.91$, $g(T_{R}) = 106.75$, $P_{\xi} = 2.1 \times 10^{-9}$ and $T_{0} = 2.37 \times 10^{-13} \GeV$, we obtain
\be{bb9} N_{*} = 59.25 - \frac{1}{2} \ln N_{*}  ~.\ee 
This is solved by $N_{*} = 57.23$. Therefore $N_{*} = 57$ is a good estimate for the case of instantaneous reheating in the conventional PQ inflation model.  

For the case of unitarity-violating PQ inflation we use \eq{e32} to eliminate $\xi$ in terms of $\lambda$,
\eq{bb5} then gives 
\be{bb11} e^{N_{*}} = \frac{c_{2} T_{0} P_{\xi}^{1/3}}{k_{*}}  \left(\frac{g(T_{0})}{g(T_{R})^{1/4}} \right)^{1/3} \frac{1}{\lambda^{1/12} N_{*}^{1/2}}   ~\ee
where 
\be{bb12} c_{2} = 2 \pi^{5/3}(12)^{\frac{1}{12}} \left(\frac{\pi^{2}}{90}\right)^{1/4} = 9.53  ~.\ee
Therefore
\be{bb13} N_{*} = 57.42 - \frac{1}{2} \ln N_{*} - \frac{1}{12} \ln \lambda  ~.\ee 
$N_{*}$ is not very sensitive to $\lambda$. For $\lambda = 0.1$ we obtain $N_{*} = 55.60$, for $\lambda = 10^{-4}$ we obtain 
$N_{*} = 56.18$, and for $\lambda = 10^{-6}$ we obtain $N_{*} = 56.55$. Thus $N_{*} = 56$ is a good estimate for the case of instantaneous reheating in the unitarity-conserving PQ inflation model when $\lambda \gae 10^{-6}$.

\end{document}